\documentclass[twocolumn,superscriptaddress,showpacs,prd,aps,amsmath,amssymb,nofootinbib]{revtex4-1}

\usepackage{graphicx}
\usepackage{longtable}
\newcommand{\beq}{\begin{equation}} 
\newcommand{\eeq}{\end{equation}} 
\newcommand{\beqn}{\begin{eqnarray}} 
\newcommand{\eeqn}{\end{eqnarray}}

\newcommand{\zD}{{\raise1.0ex\hbox{${}^{\ \circ}$}}\!\!\!\!\!D}
\newcommand{\alone}{{\raise0.5ex\hbox{${}^{\ 1}$}}\!\!\!\!\alpha}

\newcommand{\nalam}{\mathrel{\raise0.9ex\hbox{$^\lambda$}\mkern-14mu
\lower0.0ex\hbox{$\nabla$}}}

\newcommand{\zeroD}{{\raise1.0ex\hbox{${}^{\ \circ}$}}\!\!\!\!\!D}
\newcommand{\zLap}{{\raise1.0ex\hbox{${}^{\ \circ}$}}\!\!\!\!\Delta}
\newcommand{\zna}{{\raise1.0ex\hbox{${}^{\ \circ}$}}\!\!\!\!\!\nabla}
\newcommand{\zS}{{\raise1.0ex\hbox{${}^{\ \circ}$}}\!\!\!\!\!S}
\usepackage[normalem]{ulem}
\usepackage{epstopdf}
\usepackage{color}
\usepackage{soul}
\usepackage{bm}

\usepackage[]{amsmath}
\usepackage[]{amsfonts}
\usepackage[]{amssymb}
\usepackage[]{mathrsfs}
\usepackage{times}
\usepackage{hyperref}
\hypersetup{
  colorlinks=true,        % false: boxed links; true: colored links
  linkcolor=blue,         % color of internal links
  citecolor=cyan,         % color of links to bibliography
}
\def\QEQ{{%
			\setbox0\hbox{$I$}%
			\rlap{\hbox to \wd0{\hss--\hss}}\box0
		}}

\begin{document}

\title{Constraints on interquark interaction parameters with GW170817 in a binary strange star scenario}

\author{En-Ping Zhou}
%\email{}
\affiliation{State Key Laboratory of Nuclear Science and Technology and School of Physics, Peking University, Beijing 100871, China}
\affiliation{Institute for Theoretical Physics, Frankfurt am Main 60438, Germany}

\author{Xia Zhou}
\affiliation{Xinjiang Astronomical Observatory, Chinese Academy of Sciences, Urumqi, Xinjiang 830011, China}

\author{Ang Li}
\email{liang@xmu.edu.cn}
\affiliation{Department of Astronomy, Xiamen University, Xiamen, Fujian 361005, China}

\date{\today}

\begin{abstract}
The LIGO/VIRGO detection of the gravitational waves from a binary merger system, GW170817, has put a clean and strong constraint on the tidal deformability of the merging objects. From this constraint, deep insights can be obtained in compact star equation of states, which has been one of the most puzzling problems for nuclear physicists and astrophysicists. Employing one of the most widely-used quark star EOS model, we characterize the star properties by the strange quark mass ($m_s$), an effective bag constant ($B_{\rm eff}$), the perturbative QCD correction ($a_4$), as well as the gap parameter ($\Delta$) when considering quark pairing, and investigate the dependences of the tidal deformablity on them. We find that the tidal deformability is dominated by $B_{\rm eff}$, and insensitive to $m_s$, $a_4$. We discuss the correlation  between the tidal deformability and the maximum mass ($M_\mathrm{TOV}$) of a static quark star, which allows the model possibility to rule out the existence of quark stars with future gravitational wave observations and mass measurements. The current tidal deformability measurement implies $M_\mathrm{TOV} \le2.18\,M_\odot$ ($2.32\,M_\odot$ when pairing is considered) for quark stars. Combining with two-solar-mass pulsar observations, we also make constraints on the poorly known gap parameter $\Delta$ for color-flavor-locked quark matter.
\end{abstract}

\maketitle

\section{Introduction} 
\label{sec:intro}

The direct detection of the gravitational wave (GW) originating from a binary system by LIGO and VIRGO network \cite{Abbott2017}, as well as its electromagnetic (EM) counterparts detected by $\sim$ 70 astronomical observatories \cite{Abbott2017b}, has announced the birth of the long-anticipated multi-messenger astronomy era. In addition to enriching our comprehension on the central engine of short gamma ray bursts \cite{Abbott2017d,Narayan92} and the abundance of heavy elements in the Universe \cite{Abbott2017c,Eichler89}, it also contains effective information of the equation of state (EOS) of the merging objects.
Following the observation of GW170817, various works have been done on constraining the EOSs.  According to the EM counterpart, a similar upper limit of the maximum mass of a static spherically symmetric star ($M_\mathrm{TOV} \sim 2.2 M_{\odot}$) has been suggested by different groups within the neutron star (NS) model \cite{Margalit2017,Shibata2017c,Rezzolla2017,Ruiz2017,MurguiaBerthier2017,Wang2017b}. The tidal deformability measurement has also been invoked to constrain the stiffness of a generic group of NS EOSs \cite{Annala2017,Zhu2018}.

Nevertheless, the EOS of compact stars is still in lively debate, as it originates from complicated problems in non-perturbative quantum choromodynamics (QCD). Besides the conventional NS model, strange quark stars (QS) are also suggested as a possible nature of compact stars \cite{Alcock86,Itoh70}, after it was conjectured that strange quark matter (SQM) consists of up, down and strange quark could be the true ground state of strong interaction \cite{Bodmer1971,Witten84}. 
Moreover, due to the lack of information on the post-merger remnant \cite{Abbott2017,Abbott2017e}, the GW observation itself cannot exclude the possibility of a binary quark star (BQS) merger as the origin of GW170817. Additionally, attempts in understanding the EM counterparts of BQS mergers or remnant QSs have also been made \cite{Li2016,Bauswein2009,Lai2017b}. According to the estimated ejecta mass \cite{Bauswein2009} and nucleosynthesis process associated with a BQS merger \cite{Paulucci2017}, it's possible to explain the kilonova observation (AT 2017gfo) by a low opacity ejecta together with spin down power injection as suggested by \cite{Yu2017}.

QSs are quite different from NSs in many aspects, due to the self-bound nature. QSs show a different mass-radius relation compared with NSs. When supported by uniform rotation, QSs can have more enhanced mass shedding limits from their $M_\mathrm{TOV}$ values than NSs \cite{Breu2016,Bozzola2017}: $40\%$ vs. $20\%$ \cite{Li2016}. The finite surface density of QSs also requires a correction on the surface when calculating tidal deformability \cite{Damour2009,Postnikov2010}. Therefore, those constraints on NSs according to the observation of GW170817 cannot be simply applied in the scenario of BQS merger. Therefore, under the intriguing possibility that GW170817 may be from a BQS system, it will be interesting and important to study what this GW observation of tidal deformability means for QS EOS or SQM properties. 

The paper is organized as follows: In Section \ref{sec:eos}, we introduce the calculations of EOS and tidal deformability of QSs; The main results are presented in Section \ref{sec:result}, where various EOS models are extensively investigated with different choices of parameters for the strange quark mass ($m_s$), an effective bag constant ($B_{\rm eff}$), the perturbative QCD correction parameter ($a_4$), and the pairing energy gap ($\Delta$). They are all confronted with the tidal deformability ($\Lambda$) measurement from GW170817 for systematic constraints on those parameters. Finally, we summary our work and conclude in Section \ref{sec:disandconclu}.

\section{EOS models and tidal deformability}
\label{sec:eos}
Making use of the simple but widely-used MIT model \cite{Alcock86,Haensel1986}, we describe the unpaired SQM as a mixture of quarks ($u,d,s$) and electrons ($e$), allowing for the transformation due to weak interaction between quarks and leptons. The expressions for the grand canonical potential per unit volume is written as:
\beq
\Omega_{\rm free}=\sum_{i}\Omega_i^0+\frac{3}{4\pi^2}(1-a_4)\left(\frac{\mu_b}{3}\right)^4+B_\mathrm{eff}~. \label{eq:canonicalpotential1}
\eeq
In Eq.\ref{eq:canonicalpotential1},  $\Omega_i^0$ in the first term at the right hand side is the grand canonical potential for each species of the particles described as ideal Fermi gases ($i = u, d, s, e$). The second term accounts for the perturbative QCD corrections due to gluon mediated quark interactions to $\mathcal{O}(\alpha_s^2)$ \cite{Fraga2001,Alford2005,Li2017,Bhattacharyya2016}. The perturbative QCD correction parameter $a_4$ characterizes the degree of the quark interaction correction, with $a_4 = 1$ corresponding to no QCD corrections (Fermi gas approximation). $\mu_b = \mu_u + \mu_d + \mu_s$ is the baryon chemical potential, with the total baryon number density $n = (n_u + n_d + n_s)/3$. The effective bag constant ($B_\mathrm{eff}$) also includes a phenomenological representation of nonperturbative QCD effects. 

In the calculation we take $m_u=m_d=m_e=0$, and $m_s=0,\,90\,\mathrm{MeV}, 100\,\mathrm{MeV}$ \cite{Olive2014}. For $a_4$ and $B_\mathrm{eff}$, we treat them as free parameters chosen from the stability window bounded by the ``2 flavor'' line and the ``3 flavor'' line \cite{Weissenborn:2011qu}, and explore possible further limits on then with GW170817. With the constraint of ``2 flavor'' line we ensure that normal atomic nuclei shouldn't decay into non-strange quark matter. With the constraint of ``3 flavor'' line we ensure that strange quark matter would be more stable than normal nuclear matter, namely the Bodmer-Witten's conjecture \cite{Bodmer1971,Witten84}.

For a set of parameters ($m_s$, $a_4$, $B_\mathrm{eff}$), from $\Omega_{\rm free}$ in Eq.\ref{eq:canonicalpotential1} one can deduce the EOS of unpaired SQM, i.e., the pressure $p$ as a function of the energy density $\epsilon$ (the number density $n$). Then by solving the TOV equation using $p(\epsilon)$ as input, one can obtain the star's mass-radius relation $M(R)$. The obtained static maximum mass $M_\mathrm{TOV}$ should necessarily reach the present maximum mass measurement of $2.01 \pm 0.04 M_\odot$ \cite{Antoniadis2013etal}. The GW170817 observation puts independent constraints on EOS through the tidal deformability $\Lambda=(2/3)k_2/(GM/c^2R)^5$, where $k_2$ is the second Love number. $\Lambda$ describes the amount of induced mass quadrupole moment when reacting to a certain external tidal field \cite{Damour83,Damour1992b}. If a low spin prior is assumed for both stars in the binary, which is reasonable considering the magnetic braking during the binary evolution, the tidal deformability for a $1.4\,M_\odot$ star (denoted as $\Lambda$(1.4) in below) was concluded to be smaller than 800 (a more loosely constrained upper limit of 1400 is found for the high-spin prior case)~\cite{Abbott2017}.

The Love number $k_2$ measures how easily the bulk of the matter in a star is deformed by an external tidal field. Following the instructions of previous works \cite{Hinderer:2007mb,Damour2009,Yagi2013a}, we introduce a static $l=2$ perturbation to the TOV solution and solve the perturbed Einstein equation both inside and outside the star. In particular for QSs with a finite surface density, a special boundary treatment on the stellar surface has to be done to join the interior solution with the exterior \cite{Damour2009,Postnikov2010}:
\beq
y_R^\mathrm{ext}=y_R^\mathrm{int}-\frac{\epsilon_s/c^2}{M/4\pi R^3}.
\label{eq:surfcorr}
\eeq
In Eq.\ref{eq:surfcorr}, $y=rH^\prime/H$ is a variable relating to $k_2$ with a complicated algebraic expression and $H$ (as a function of $r$) is the perturbation introduced into the metric and satisfies a second order differential equation with metric and fluid variables. $\epsilon_s$ is the surface energy density.

\section{Results}
\label{sec:result}
\subsection{Dependence of $\Lambda$ on $m_s$}

\begin{figure}
	\begin{center}
		\includegraphics[height=70mm]{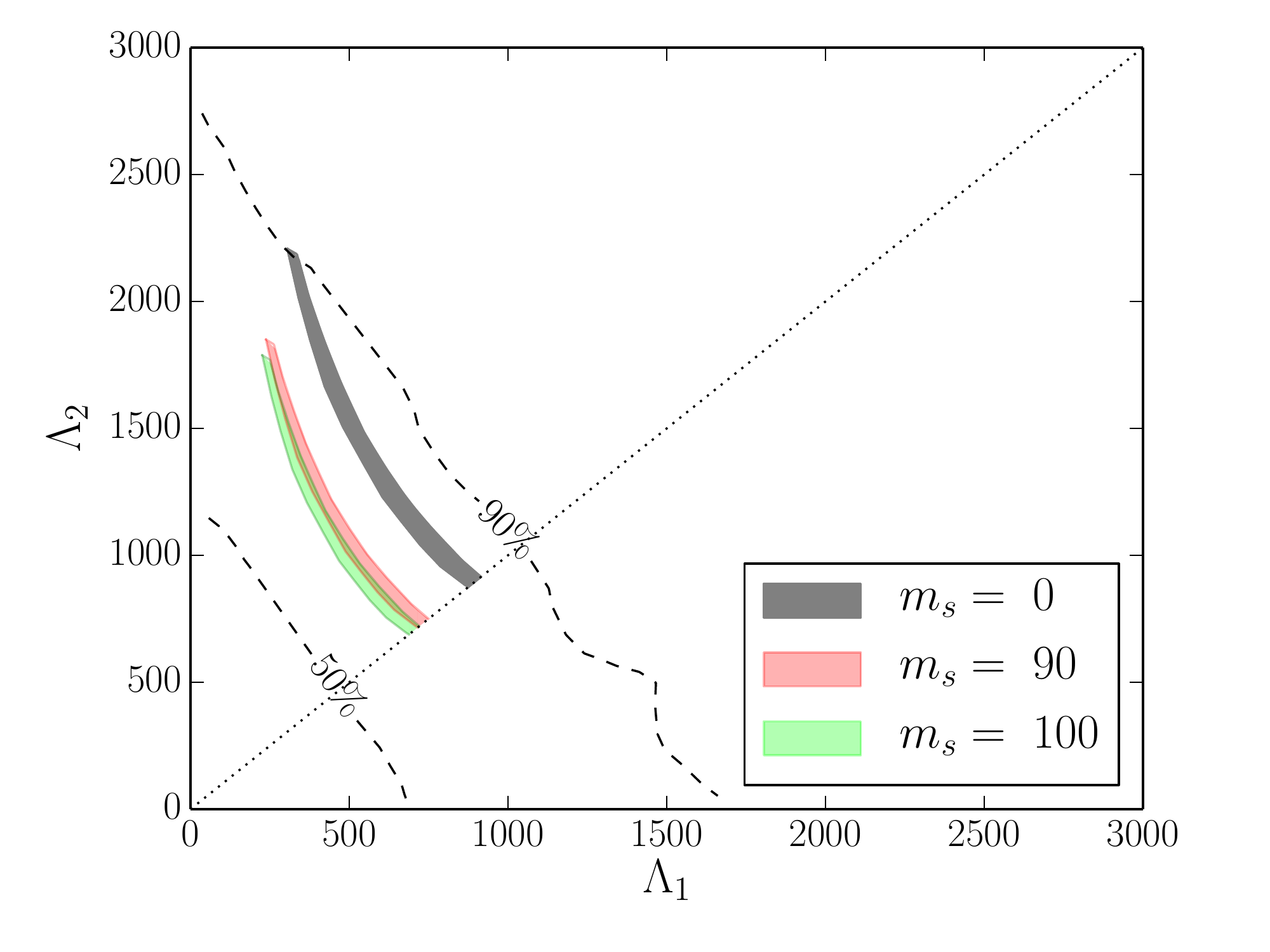}
	\end{center}
	\caption{Tidal deformability ($\Lambda_1$, $\Lambda_2$) for $m_s = 0$, $90\,\mathrm{MeV}$, $100\,\mathrm{MeV}$, with fixed $a_4=0.61$ and $B_\mathrm{eff}^{1/4}=138\,\mathrm{MeV}$. $\Lambda_1$ and $\Lambda_2$ are calculated by employing the 90$\%$ most probable fraction of component masses $M_1$ and $M_2$ for GW170817. They are compared with the observation \cite{Abbott2017}: 2 dashed lines enclose 50$\%$ and 90$\%$ of the of the probability density, respectively, in the low spin prior case.}
	\label{fig:l1l2}
\end{figure}

\begin{table}
	\begin{tabular}{ccccccccccc}
		\hline
		$m_s\,[\mathrm{MeV}]$  & $n_c\,[\mathrm{fm}^{-3}]$ 
		& $R\,[\mathrm{km}]$ & $M/R$ & $k_2$ & $\Lambda$  \\
		\hline
		0   & 0.327  & 11.814 & 0.17499 & 0.19510 & 792.8  \\
		90  & 0.355  & 11.478 & 0.18016 & 0.18357 & 644.9  \\
		100 & 0.361  & 11.415 & 0.18115 & 0.18133 & 619.7  \\
		\hline
	\end{tabular}
	\caption{Properties of a $1.4\,M_\odot$ QS, including the central number density $n_c$, the radius $R$, the compactness $M/R$, the Love number $k_2$ and the tidal deformability $\Lambda$, for different $m_s$ with fixed $a_4=0.61$ and $B_\mathrm{eff}^{1/4}=138\,\mathrm{MeV}$.}
	\label{tab:res_squarkmass}
\end{table}

$m_s$ has been well-constrained, for a recent result of $95\pm5\,\mathrm{MeV}$ \cite{Olive2014}. The tiny uncertainty of $m_s$ allow us to verify easily whether there is strong dependence of $\Lambda$ on it. For this purpose, we have applied three models with same usual values of $a_4$ (0.61) and $B_\mathrm{eff}^{1/4}$ (138\,$\mathrm{MeV}$) but different $m_s$ (0, 90\,MeV and 100\,MeV). The results are shown in Fig.\ref{fig:l1l2} and Table.\ref{tab:res_squarkmass}. Three QS EOSs are all consistent with the observation of GW170817, justifying the possibility of BQS merger despite those differences between QSs and NSs. The massless case gives $\Lambda(1.4)=791.4$, still managing to be inside the boundary of GW170817 constraint.

The effect of bringing in finite strange quark mass is to soften the EOS. Consequently, the TOV maximum mass decreases from 2.217 $M_\odot$ in the massless case to 2.101 $M_\odot$ (2.079 $M_\odot$) for $m_s = 90\,\mathrm{MeV}$ ($m_s = 100\,\mathrm{MeV}$). Also for a fixed gravitational mass of 1.4\,$M_\odot$, with increasing $m_s$, the central density increases, the star radius decreases, resulting an increasing compactness and a decreasing Love number. Those lead to a decreasing tidal deformability. However, the differences between $\Lambda$(1.4) for the two finite $m_s$ models are negligible. That is, $\Lambda$(1.4) only weakly depends on $m_s$.

\subsection{Dependence of $\Lambda$ on $a_4$ and $B_\mathrm{eff}$}

\begin{figure}
	\begin{center}
		\includegraphics[height=70mm]{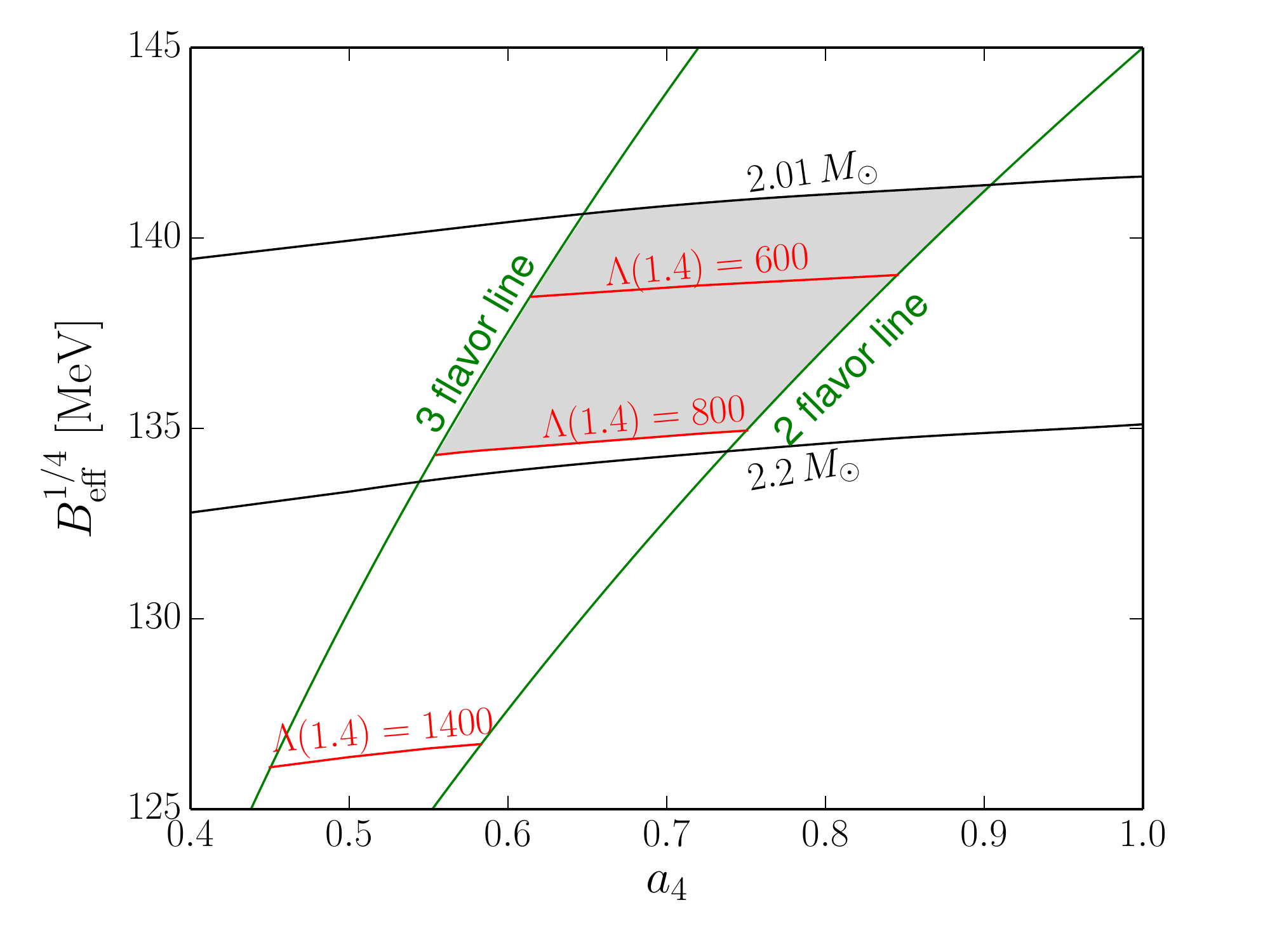}
	\end{center}
	\caption{Constraints on $a_4$ and $B_\mathrm{eff}$ with $m_s=100\,\mathrm{MeV}$. The curves of measurement of $\Lambda$ according to GW170817 for both the low-spin prior and the high-spin prior (two lower red lines) are demonstrated, along with those of mass measurement of pulsars (top black line) and stability condition (green lines). The $\Lambda(1.4)=600$ line and the $M_\mathrm{TOV}=2.2\,M_\odot$ line indicate possible future observational constraints. We also illustrate the allowed parameter space jointly constrained by red/black/green lines for the low-spin prior.}
	\label{fig:B_a4}
\end{figure}

\begin{table}
	\begin{tabular}{ccccccccccc}
		\hline
		$a_4$ & $B_{\rm eff}^{1/4}\,[\mathrm{MeV}]$ 
		& $R\,[\mathrm{km}]$ & $M/R$ & $k_2$ & $\Lambda$  \\
		\hline
		0.61 & 133 & 12.046 & 0.17166 & 0.19973 & 893.4 \\
		0.61 & 136 & 11.662 & 0.17731 & 0.18865 & 717.7 \\
		0.61 & 138 & 11.415 & 0.18115 & 0.18133 & 619.7 \\
		0.72 & 138 & 11.453 & 0.18055 & 0.18262 & 634.5 \\
		0.83 & 138 & 11.482 & 0.18008 & 0.18367 & 646.6 \\
		
		\hline
	\end{tabular}
\caption{Properties of a $1.4\,M_\odot$ QS, including the radius $R$, the compactness $M/R$, the Love number $k_2$ and the tidal deformability $\Lambda$, for various choices of $a_4$ and $B_\mathrm{eff}$ with fixed $m_s = 100\,\mathrm{MeV}$.}
\label{tab:res_a4B}
\end{table}

Since the dependence of $\Lambda$ on $m_s$ is very weak, we fix strange quark mass as 100 MeV in the following. Selected solutions with several sets of ($a_4$, $B_\mathrm{eff}$) are presented in Table.\ref{tab:res_a4B} and interpolations have been done to produce a series of constant $M_\mathrm{TOV}$ lines and $\Lambda$(1.4) lines in Fig.\ref{fig:B_a4}. It is clearly shown that softer EOSs are more compact for a given mass, and less likely to be tidally deformed \cite{Hinderer09,Postnikov2010}.

Both $M_\mathrm{TOV}$ and $\Lambda$(1.4) increases with increasing $a_4$. Namely, the perturbative QCD corrections soften the EOS as well as the finite strange quark mass, although the softening effect is quite modest and the constant $M_\mathrm{TOV}$ lines in Fig.\ref{fig:B_a4} is close to horizontal ones. Unlike the weak dependence of $\Lambda$(1.4) on $m_s$ and $a_4$, the effective bag constant is dominating for determining the tidal deformability and maximum mass. Consequently, a rather proper constraint can be set on $B_\mathrm{eff}$ with GW170817. 

Combining the GW170817 constraint on $\Lambda$(1.4), the two-solar-mass constraint on $M_\mathrm{TOV}$ and the stability window for quark matter, we find that the QS model parameters to be compatible with $B_\mathrm{eff}^{1/4}\in(134.1,141.4)$\,MeV and $a_4\in(0.56,0.91)$ for the low-spin prior. The parameter space for the high spin prior case is relatively larger, i.e. $B_\mathrm{eff}^{1/4}\in(126.1,141.4)$\,MeV and $a_4\in(0.45,0.91)$, as the tidal deformability is more loosely bound by observation in this case. Further limits can be added once more observations are made in the future, as indicated Fig.\ref{fig:B_a4} with the $\Lambda(1.4)=600$ line and the $M_\mathrm{TOV}=2.2\,M_\odot$ line.

\subsection{Correlation between $\Lambda$ and $M_\mathrm{TOV}$}

\begin{figure}
\begin{center}
\includegraphics[height=70mm]{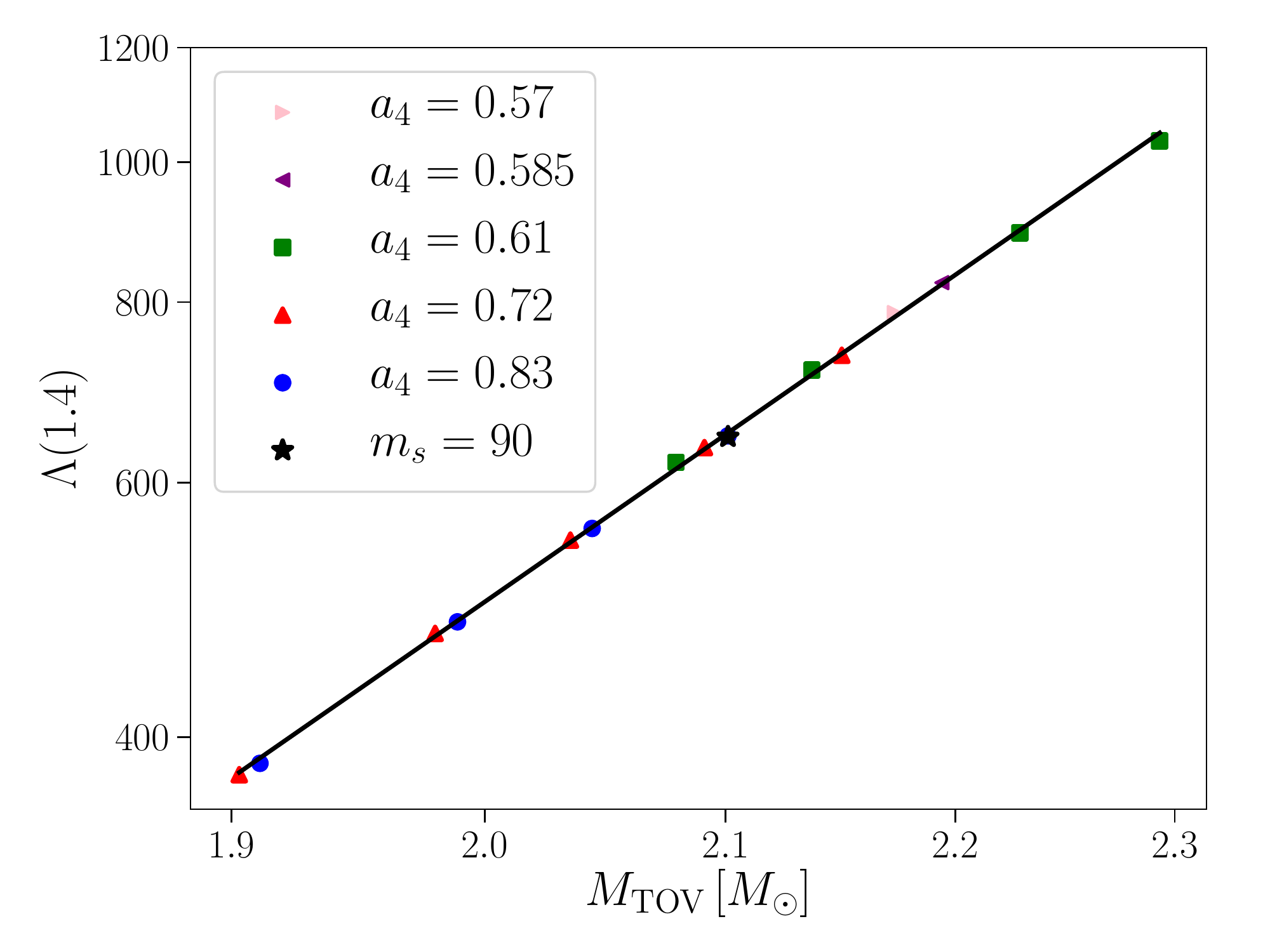}
\end{center}
\caption{Correlation between $\Lambda$(1.4) and $M_\mathrm{TOV}$ for various $a_4$ and $B_\mathrm{eff}$, for $m_s = 100\,\mathrm{MeV}$ (color symbols) or $m_s = 90\,\mathrm{MeV}$ (black symbols). The solid line indicates our linear fit of Eq.\ref{eq:mlambdafit}. Symbols with different colors and shapes lying in the same line demonstrate that the fitting formula is independent of $a_4$ and $m_s$.}
\label{fig:M_lambda}
\end{figure}

Since both $M_\mathrm{TOV}$ and $\Lambda$(1.4) indicate the stiffness of QS EOSs, particularly, their dependences on both $B_\mathrm{eff}$ and $a_4$ are almost identical, it will be useful to present a straightforward relation between them. As can be seen from Fig.\ref{fig:M_lambda}, a strong linear dependence is found between $M_\mathrm{TOV}$ and $\Lambda$(1.4) in logarithm scale, which is independent of $a_4$. We fit the data in the following form:
\beq
\Lambda(1.4) = 510.058\times \left(\frac{M_\mathrm{TOV}}{2.01\,M_\odot}\right)^{5.457},
\label{eq:mlambdafit}
\eeq
with coefficient of determination $R^2 = 0.9996$. We also check that the $m_s=90\,\mathrm{MeV}$ case can also satisfy the same fitted formula.

A direct application of the strong $M_\mathrm{TOV}$-$\Lambda$(1.4) correlation is to set a constraint on QS EOS directly from a tidal observation. For example, the corresponding $M_\mathrm{TOV}$ for $\Lambda(1.4)=800$ is 2.18\,$M_\odot$. Although the physical picture is different in many aspects, we coincidentally end up with a similar upper limit of $M_\mathrm{TOV}$ for QSs as NSs \cite{Margalit2017,Shibata2017c,Rezzolla2017,Ruiz2017} from the observation of GW170817. 

On the other hand, it is also possible to find the minimum possible $\Lambda$(1.4) value with the current two-solar-mass constraint, which is 510.1. Once future GW observations find a smaller upper limit, it will be impossible for a QS EOS to accommodate in the present model, unless other quark-matter phases is included such as quark pairing. We mention here that $\Lambda$(1.4) for the NS EOS of APR4 (consists of $n,p,e$, and $\mu$ \cite{Akmal1998a}) is 255.8.

\subsection{Dependence of $\Lambda$ on $\Delta$}

\begin{figure}
	\begin{center}
		\includegraphics[height=70mm]{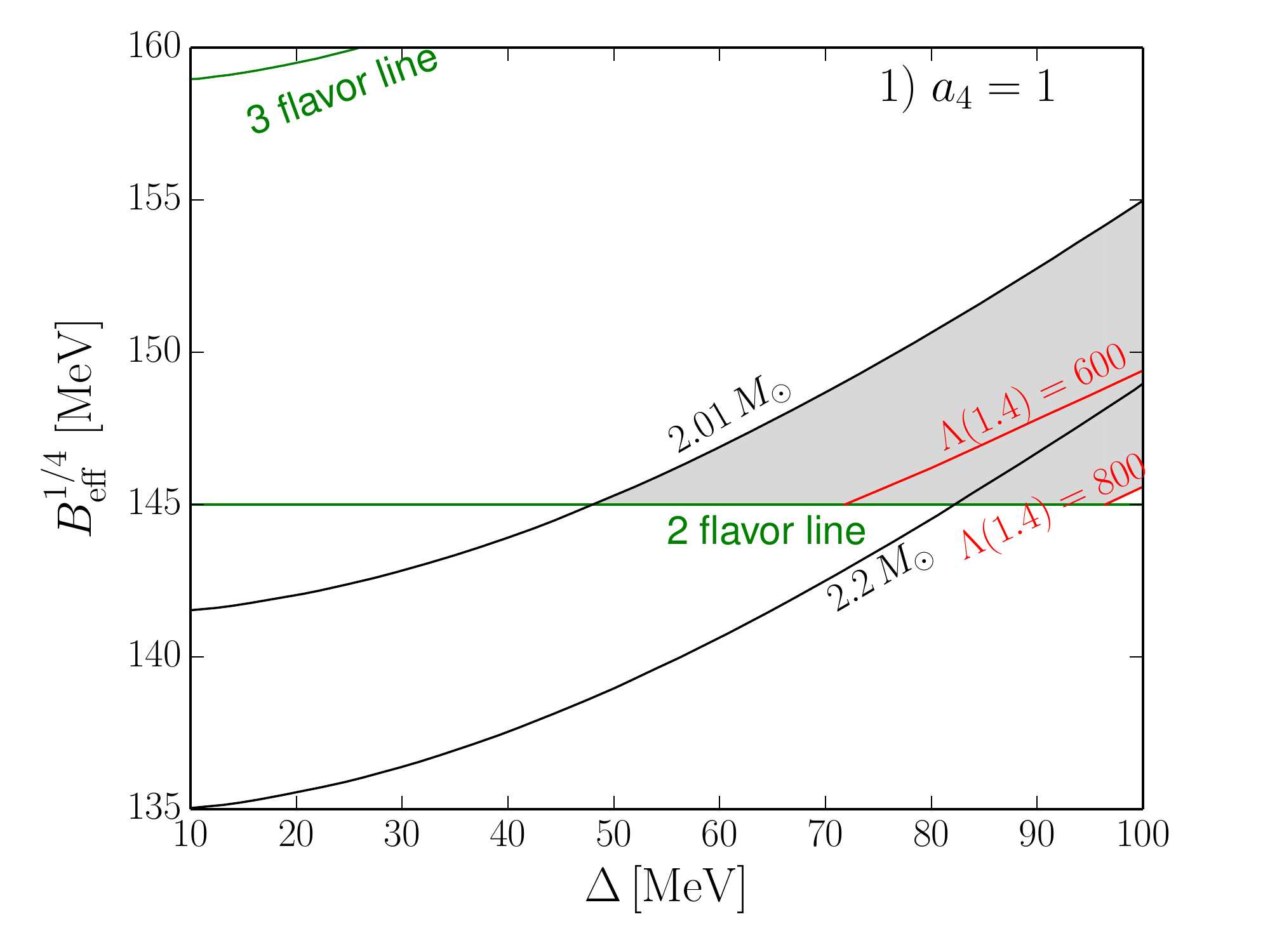}
		\includegraphics[height=70mm]{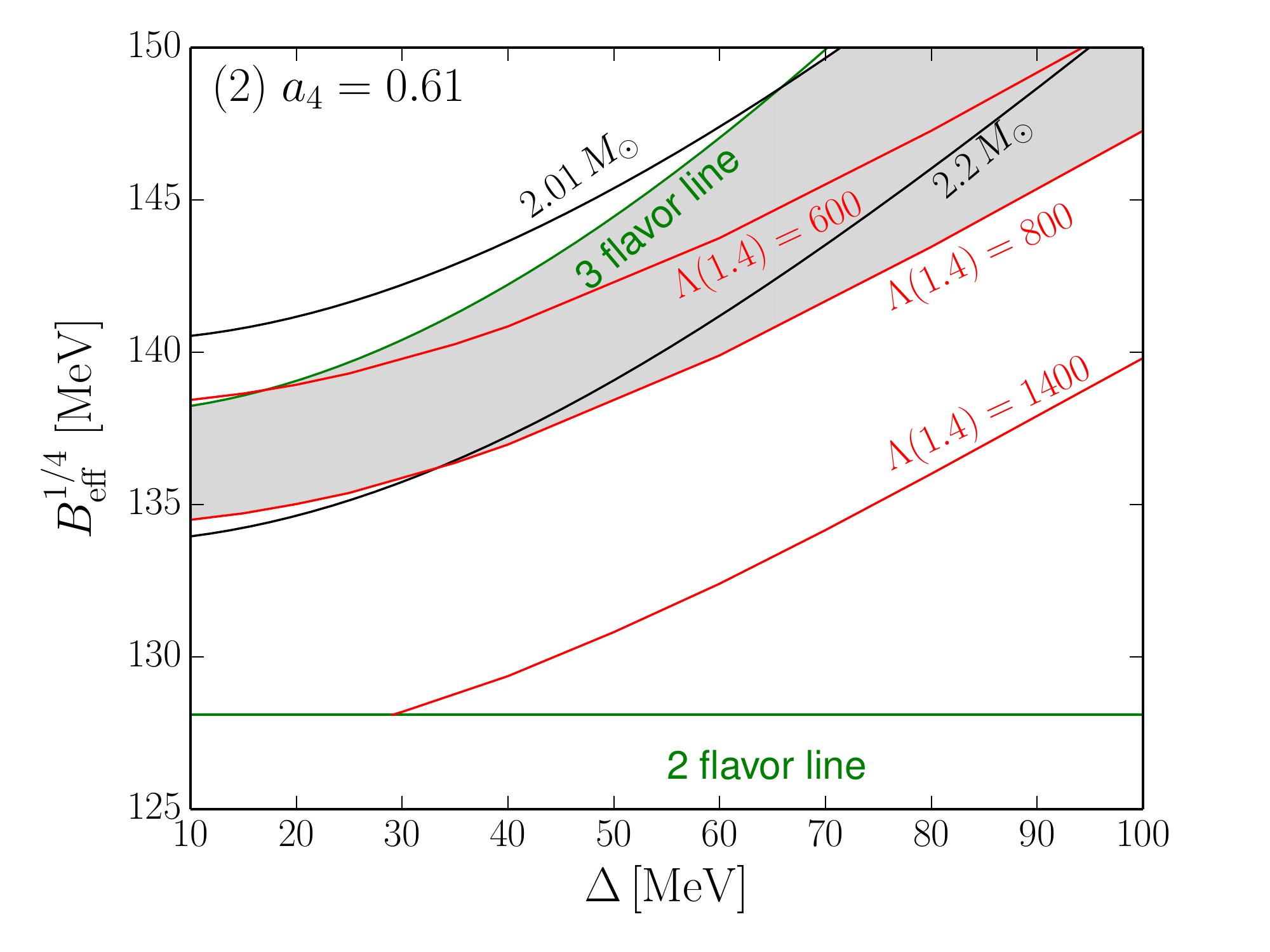}
	\end{center}
	\caption{Constraints on the pairing gap $\Delta$ with $m_s=100\,\mathrm{MeV}$, for both $a_4 = 1$ (upper panel) and $a_4 = 0.61$ (lower panel). $a_4 = 1$ corresponds to no perturbative QCD correction, and $a_4 = 0.61$ is close to the previous calculated result with different choice of the renormalization scale \cite{Fraga2001}. Other notions applied are the same with Fig.\ref{fig:B_a4}.}
	\label{fig:B_Delta}
\end{figure}

If quarks form Cooper pairs and the SQM is in the color-flavor-locked (CFL) phase, an additional term corresponding to the energy of the diquark condensate has to be added:
\beq
\Omega_{\rm CFL} = \Omega_{\rm free} - \frac{3}{\pi^2}\Delta^2\mu_b^2~.
\label{eq:canonicalpotential2}
\eeq
$\Delta$ in Eq.\ref{eq:canonicalpotential2} is the pairing energy gap for the phase, lacking an accurate calculation within a typical range ($0-100$ MeV \cite{Alford1999,Rajagopal2001,Lugones2002,Rischke2004}, possibly up to $150\,\mathrm{MeV}$ \cite{Flores2017}). Starting with this grand canonical potential, we here explore the possibility of a new constraint for $\Delta$ with GW170817. There may be other quark-matter phases, such as two-flavor color-superconducting (2SC) phase (e.g., \cite{Alford2017b}), but we leave a study of these to future work.

The result is shown in Fig.\ref{fig:B_Delta}. In the ideal $a_4=1$ case, the $\Lambda(1.4)\leq800$ constraint is found to be consistent with the previous upper value of $\Delta \ge 100$ MeV, except for when $B_\mathrm{eff}$ choice is close to the ``2 flavor line'' condition. The two-solar-mass constraint bounds the lower limit of $\Delta$ at the order of $\sim50\,$MeV. In the more realistic $a_4=0.61$ case, the curves of constant $M_\mathrm{TOV}$ and $\Lambda$(1.4) are very close with the $a_4=1$ case, indicating that the EOS $p(\epsilon)$ only weakly depends on $a_4$, similar as in the unpaired case of Fig.~\ref{fig:B_a4}. The constraints from nuclear physics (i.e., "2 flavor line" and "3 flavor line" conditions) do change significantly, which can also be inferred from the unpaired case (c.f. Fig. \ref{fig:B_a4}): larger effective bag constant is required to bound the quark matter for larger $a_4$. Consequently in the $a_4=0.61$ case, no new lower limit is found for the gap parameter $\Delta$, for both the low-spin prior and the high-spin prior. The present calculations can reconcile with the previous range for $\Delta$.

\section{Conclusion}
\label{sec:disandconclu}

The properties of QSs have been systematically studied using one of the most widely-used QS EOS model, for the purpose of constraining the QS EOS and interquark interaction parameters with the tidal deformability of GW170817. We find that a QS should have $M_\mathrm{TOV}\le2.18\,M_\odot$ ($2.32\,M_\odot$ when considering CFL phase). In particular, a power law relation between $M_\mathrm{TOV}$ and $\Lambda$(1.4) is newly found and fitted for normal SQM. 

We also demonstrate that finite $m_s$ play only minor role, while it is $B_\mathrm{eff}$ dominating the EOS stiffness, compared with a weak influence of $a_4$. Combining GW and pulsar observations, $B_\mathrm{eff}^{1/4}$ is limited in a range of (134.1,141.4)\,$\mathrm{MeV}$, and $a_4$ between 0.56 to 0.91 for a realistic low-spin prior and a relatively larger parameter space is found ($B_\mathrm{eff}^{1/4}\in(126.1,141.4)$\,MeV and $a_4\in(0.45,0.91)$) for the high-spin prior case. Furthermore, the $\Lambda(1.4)$ constraint of GW170817 is found to be consistent with its previous upper value ($ \ge 100\,\mathrm{MeV}$).

In the present study it is shown that for both paired and unpaired SQM, GW170817 has the possibility of originating from a BQS merger, and the GW observation of tidal deformability can translate into constraints for high-density QS EOSs. By combining with the constraints from massive ($>2.0M_{\odot}$) pulsars, the model parameters of QS EOSs are ready to be more tightly defined by future observations. If future GW and pulsar observations could not be reconciled with each other, QSs might be excluded as possible type of compact stars using the present model.

\acknowledgments
We thank Dr. A. Tsokaros and C. Pan for useful discussion and helps in the numerical calculation. E. Z. is grateful for China Scholarship Council for supporting the joint training in Frankfurt. The work was supported by the National Natural Science Foundation of China (Nos. U1431107 and 11373006), and the West Light Foundation of Chinese Academy of Sciences (No. ZD201302).

\bibliographystyle{apsrev4-1}

\end{document}